\documentclass[aps,pre,twocolumn,superscriptaddress]{revtex4-1}
\usepackage{amsmath}
\usepackage{amssymb}
\usepackage{graphicx}
\usepackage{subfigure}

\begin{document}

%\preprint{}

\title{Dynamical heterogeneities in liquid and glass originate from medium-range order}


\author{Charles K. C. Lieou}
\email[]{clieou@utk.edu}
%\thanks{}
\affiliation{Department of Materials Science and Engineering, University of Tennessee, Knoxville, TN 37996, USA}
\author{Takeshi Egami}
\affiliation{Department of Materials Science and Engineering, University of Tennessee, Knoxville, TN 37996, USA}
\affiliation{Department of Physics and Astronomy, University of Tennessee, Knoxville, TN 37996, USA}
\affiliation{Materials Science and Technology Division, Oak Ridge National Laboratory, Oak Ridge, TN 37831, USA}

\date{\today}

\begin{abstract}
Slow relaxation and plastic deformation in disordered materials such as metallic glasses and supercooled liquids occur at dynamical heterogeneities, or neighboring particles that rearrange in a correlated, cooperative manner. Dynamical heterogeneities have historically been described by a four-point, time-dependent density correlation function $\chi_4 (r, t)$. In this paper, we posit that $\chi_4 (r, t)$ contains essentially the same information about medium-range order as the Van Hove correlation function $G(r, t)$. In other words, medium-range order is the origin of spatially correlated regions of cooperative particle motion. The Van Hove function is the preferred tool for describing dynamical heterogeneities than the four-point function, for which the physical meaning is less transparent.
\end{abstract}

%\keywords{}

\maketitle


\section{Introduction}\label{sec:1}

Structural relaxation and plastic deformation in disordered materials, such as supercooled liquids, metallic glasses, and granular materials, occur via the correlated motion of atoms or constituent particles \cite{adam_1965,argon_1979,falk_1998,falk_2011}. Such collective motions characterize dynamical heterogeneities \cite{kob_1997, donati_1998, zhang_2005, widmer_2006, biroli_2008, fan_2014, qiao_2017, wang_2019}. To elucidate the correlation between particles, one may introduce a four-point function $\chi (r, t)$, which measures the spatial correlation between two different positions at two different times \cite{glotzer_2000}. Ref.~\cite{lacevic_2003} is among the most extensive studies of the four-point function. There, the authors posit that the four-point function describes dynamical heterogeneities and exhibit a correlation length that grows with decreasing temperature in a glass-forming liquid, indicative of the correlated motion of atoms whose cluster size increases with cooling. 

From the four-point function $\chi (r, t)$, one may define a four-point susceptibility, or global four-point function $\chi_4 (t)$ from which one can compute the size of a typical dynamical heterogeneity, via a simple counting argument described in \cite{abate_2007}. Indeed, if one defines the time-dependent order parameter $Q(t)$ by
\begin{equation}
 Q(t) = \dfrac{1}{N} \sum_{i = 1}^N w_i (t) ,
\end{equation}
where $w_i(t)$ is the threshold function
\begin{equation}\label{eq:w_i}
 w_i (t) = 
 \begin{cases}
  1 , \quad \text{if $| \mathbf{r}_i (t) - \mathbf{r}_i (0) | < d$} ;  \\
  0 , \quad \text{if $| \mathbf{r}_i (t) - \mathbf{r}_i (0) | \geq d$},
 \end{cases}
\end{equation}
for each of the $N$ atoms $i$ in the system, i.e., $w_i(t)$ equals unity if the displacement of atom $i$ over time $t$ does not exceed some threshold $d$, and zero otherwise, then one can naturally define the global four-point function
\begin{equation}\label{eq:chi4_def}
 \chi_4 (t) = N \left( \left\langle Q (t) ^2 \right \rangle - \left\langle Q(t) \right\rangle^2 \right) .
\end{equation}
It can be shown easily \cite{abate_2007} that the typical number of particles in a dynamical heterogeneity is given by
\begin{equation}
 n = \dfrac{\chi_4 (t^*)}{1 - Q(t^*)},
\end{equation}
where $t^*$ is the value of $t$ at which $\chi_4 (t)$ attains its peak value. The reader may immediately appreciate that the classification of whether a particle belongs to a dynamical heterogeneity depends on the threshold displacement $d$.

However, the precise physical meaning of the four-point correlation function $\chi (r, t)$ itself remains elusive. It is recently proposed \cite{ryu_2019,ryu_2021,egami_2021,lieou_2022a} that the occurrence of dynamical heterogeneities is associated with medium-range order (MRO) in glass-forming materials. The MRO in a glass-forming liquid is observed as oscillations beyond the nearest neighbor peak in the Van Hove function, $G(r, t)$ \cite{vanhove_1954}, or the pair distribution function (PDF), $g(r)$ \cite{egelstaff_1991}. The oscillations decay with a characteristic coherence length, $\xi$, which depicts the range of cooperativity. Both the Van Hove function and PDF are defined in a much more straightforward manner, with rather transparent physical meanings. They can be determined by experiment through x-ray or neutron scattering \cite{egami_2020b,warren_1969}, and $\xi$ is directly related to viscosity \cite{ryu_2019} and liquid fragility \cite{ryu_2020}. This motivates the present work, in which we investigate the relationship between $\chi (r, t)$ and the Van Hove function $G(r, t)$, and examine whether $\chi(r, t)$ carries information about MRO.

The rest of the paper is structured as follows. In Sec. \ref{sec:2} we define the four-point function. In contrast to the approach taken by Ref. \cite{lacevic_2003}, we use the \textit{global} four-point susceptibility in Eq.~\eqref{eq:chi4_def} to motivate the local, position-dependent four-point function. We compare the four-point function to $g(r)$ and $G(r, t)$ in Sec.~\ref{sec:3}, and posit that MRO is the origin of dynamical heterogeneities. We conclude the paper with some additional remarks and insights in Sec.~\ref{sec:4}.

\section{The four-point function}
\label{sec:2}

Motivated by the definition in Eq.~\eqref{eq:chi4_def}, we define the \textit{local}, position-dependent four-point function
\begin{equation}\label{eq:chi_rr}
 \chi_4 (\mathbf{r}', \mathbf{r}'' , t) \equiv \left\langle Q ( \mathbf{r}', t ) Q (\mathbf{r}'', t ) \right\rangle - \left\langle Q (\mathbf{r}', t) \right\rangle \left\langle Q(\mathbf{r}'', t) \right\rangle, ~~~~~
\end{equation}
where
\begin{equation}
 Q (\mathbf{r}', t) = \sum_{i=1}^N Q_i (t) \delta (\mathbf{r}' - \mathbf{r}_i(0)) ,
\end{equation}
and we define the quantity $Q_i$ for atom $i$ in one the following ways:
\begin{eqnarray}
 \label{eq:Q1} Q_i (t) &\equiv& Q_{N,i} (t) = \sum_j \dfrac{1}{z_i + 1} w_j (t) ; \quad \text{or} \\
 \label{eq:Q2} Q_i (t) &\equiv& Q_{S,i} (t) = w_i (t) .
\end{eqnarray}
In Eq.~\eqref{eq:Q1} the sum is over all $z_i$ neighbors (within some cutoff distance) of each atom $i$ plus itself; $w_i (t)$ is the thresholding function defined above in Eq.~\eqref{eq:w_i}. To convert this into a radially symmetric function that depends only on the relative position $\mathbf{r} \equiv \mathbf{r}' - \mathbf{r} ''$, integrate Eq.~\eqref{eq:chi_rr} to get
\begin{eqnarray}
 \nonumber & & \int d \mathbf{r}' \, d \mathbf{r}'' \chi_4 (\mathbf{r}', \mathbf{r}'', t) \\ \nonumber &=& \sum_i \sum_j \int d \mathbf{r} \left\langle Q_i(t) Q_j(t) \delta (\mathbf{r} - \mathbf{r}_i + \mathbf{r}_j ) \right\rangle \\ & & - \sum_i \sum_j \left\langle Q_i(t) \right\rangle \left\langle Q_j(t) \right\rangle . ~~~~~
\end{eqnarray}
In the above and what follows, we use $\mathbf{r}_i$ as a shorthand for $\mathbf{r}_i (0)$, the position of particle $i$ at the initial time $t = 0$. To get a dimensionless, intensive quantity independent of system size, and anticipating the connection to the PDF $g(r) = g(\mathbf{r})$, multiply this through by $V / N^2$, where $V$ is the extensive volume of the system. Then we see immediately that the appropriate definition of the $\mathbf{r}$-dependent four-point function is
\begin{eqnarray}\label{eq:chi_r}
 \nonumber \chi (\mathbf{r}, t) &=& \dfrac{V}{N^2} \sum_{i} \sum_{j} \left\langle Q_i (t) Q_j(t) \delta (\mathbf{r} - \mathbf{r}_i + \mathbf{r}_j) \right\rangle \\ & & - \sum_i \sum_j \left\langle \dfrac{Q_i(t)}{N} \right\rangle \left\langle \dfrac{Q_j(t)}{N} \right\rangle .
\end{eqnarray}
One sees that if we take $Q_i(t) = w_i(t)$ in Eq.~\eqref{eq:chi_r}, as defined in Eq.~\eqref{eq:Q2} above, we arrive at a function that is identical to what is called $g_4(r)$ in \cite{lacevic_2003}; we shall call this $\chi_S (\mathbf{r}, t)$. If we choose $Q_i(t)$ according to Eq.~\eqref{eq:Q1}, we call the resultant four-point function $\chi_N (\mathbf{r}, t)$. Also, we see that in the $t \rightarrow 0$ limit,
\begin{equation}\label{eq:chi_r_t0}
 \chi(\mathbf{r}, 0) = \dfrac{V}{N^2} \sum_i \sum_j \left\langle \delta( \mathbf{r} - \mathbf{r}_i + \mathbf{r}_j ) \right\rangle - 1 = g(\mathbf{r}) - 1.
\end{equation}
is indeed related to the PDF, because $Q_i (t \rightarrow 0) = 1$ -- as all atoms stay within their cages in this limit. This holds for both $\chi_S(\mathbf{r}, t)$ and $\chi_N (\mathbf{r}, t)$. It is also obvious that one can rewrite Eq.~\eqref{eq:chi_r} to reflect the isotropy:
\begin{eqnarray}
 \nonumber \chi (r, t) &=& \dfrac{V}{N^2} \sum_i \sum_j \dfrac{\left\langle Q_i Q_j \delta(r - | \mathbf{r}_i - \mathbf{r}_j|)\right\rangle}{4 \pi r^2} \\ & & - \sum_i \sum_j \left\langle \dfrac{Q_i(t)}{N} \right\rangle \left\langle \dfrac{Q_j(t)}{N} \right\rangle .
\end{eqnarray}

\section{Medium-range order: simulations and results}
\label{sec:3}

We performed molecular dynamics simulations using the Large Scale Massively Parallel Simulator (LAMMPS) \cite{plimpton_1995}. The supercooled Fe sample was prepared with a melt-quench method, starting with $N = 16000$ Fe atoms, interacting with the modified Johnson potential and in a bcc lattice with a cubic side length of 58.86 \r{A}, melted at 5000 K, with periodic boundary conditions. The sample was then supercooled at a rate of $10^3$ K/ps in an NVT ensemble and equilibrated over 1 $\mu$s, in intervals of 100 K, down to a temperature of 1500 K, which is well above the glass transition temperature ($\sim 950$ K). Then we hold the temperature fixed at 1500 K and compute the four-point functions $\chi_N (r, t)$ and $\chi_S (r, t)$, with thresholds (cf.~Eq.~\eqref{eq:w_i}) $d = a/2$, $a$, and $2 a$, where $a = 0.824$ \r{A} is the mean-square displacement over $\Delta t = 1$ ps. We also compute the Van Hove function
\begin{equation}
 G(r, t) = \dfrac{V}{N^2} \sum_i \sum_j \dfrac{\left\langle \delta(r - |\mathbf{r}_i (t) - \mathbf{r}_j (0)|)\right\rangle}{4 \pi r^2} ,
\end{equation}
and compare its behavior to that of the four-point functions.

\begin{figure}[t]
\begin{center}
\includegraphics[scale=0.5]{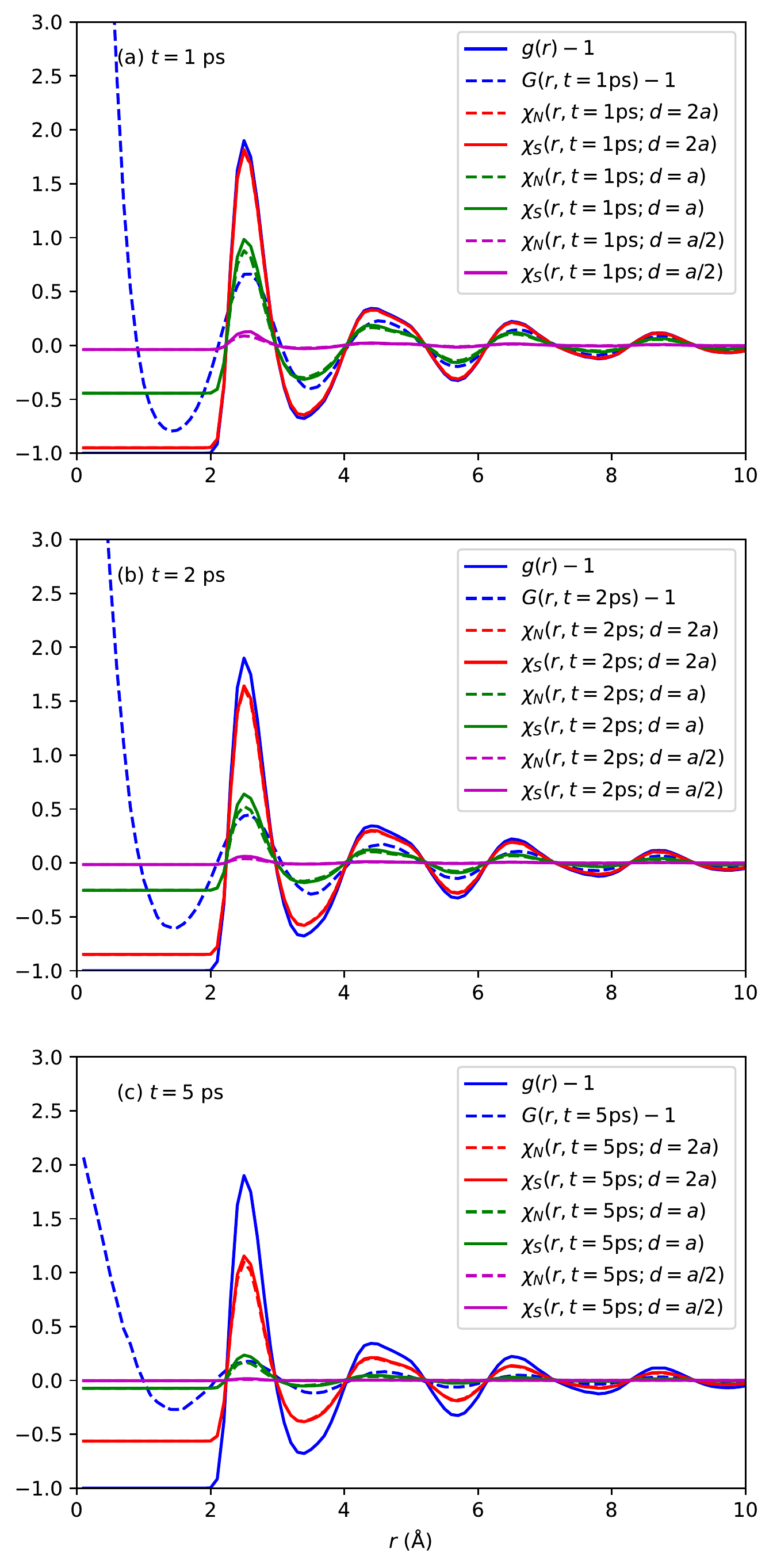}
\caption{\label{fig:coherence_all}The pair-distribution function (PDF) $g(r)$, the Van Hove correlation function $G(r, t)$, and the four-point functions $\chi_N (r, t)$ and $\chi_S (r, t)$ with thresholds $d = a/2$, $a$, and $2 a$, where $a$ is the mean-square displacement over 1 ps, for various times (a) $t = 1$ ps, (b) $t = 2$ ps, and (c) $t = 5$ ps.}
\end{center}
\end{figure}

Figure \ref{fig:coherence_all} is a direct comparison of the PDF $g(r)$, the Van Hove function $G(r, t)$, and the four-point functions $\chi_N(r, t)$ and $\chi_S (r, t)$, for various displacement thresholds $d$, at times $t = 1$, 2, and 5 ps. It becomes immediately apparent that the four-point functions decay from $\chi_{N,S} (r, t = 0) = g(r)$ and that their peak positions are unchanged with time. The decay is more pronounced with decreasing threshold distance $d$; this is easily understood, for a reduction of $d$ means that more atoms are regarded to have been displaced outside of their cage of size $d$ over the same amount of time, decreasing the correlation. The quantities $\chi_N (r, t)$ and $\chi_S (r, t)$ behave almost identically, at all times and for all threshold distances. In particular, the choice $d = a$ gives rise to a four-point function which roughly traces the behavior of the Van Hove function $G(r, t)$ at all times; because $a$ is the mean-square displacement over 1 ps, this suggest that the characteristic time scale of liquid Fe at 1500 K is indeed roughly equal to 1 ps. In any case, the observation that the behaviors of $\chi_N (r, t)$ and $\chi_S (r, t)$ mirror that of $G(r, t)$ suggests that these quantities carry the same information about medium-range order in the glass-forming liquid. As such, we identify MRO to be the origin of dynamical heterogeneities in glass-forming liquids.

\begin{figure}[ht]
\begin{center}
\includegraphics[scale=0.5]{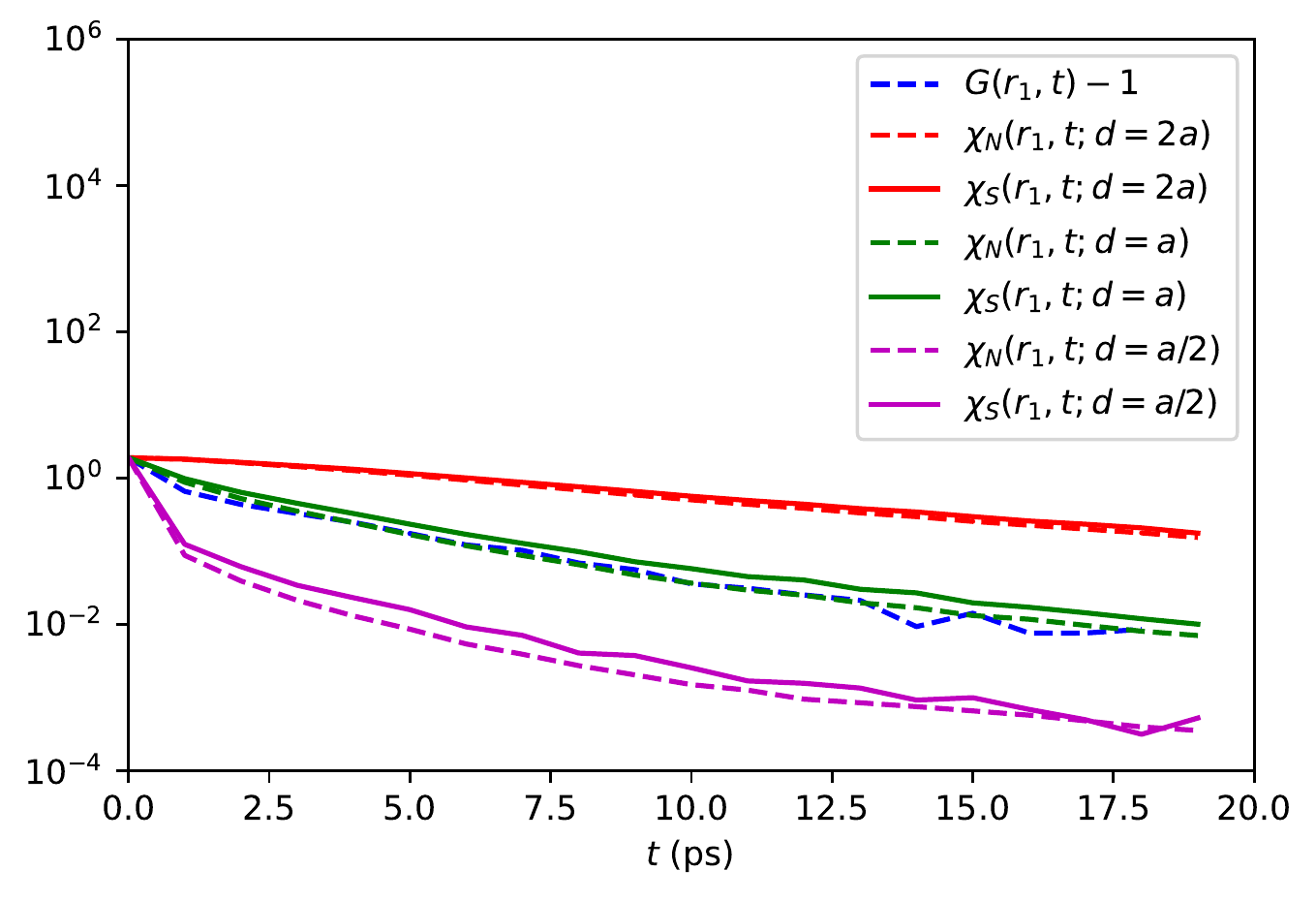}
\caption{\label{fig:coherence_fp}Decay of the first peak of the Van Hove function $G(r, t)$ and the four-point functions $\chi_N(r, t)$ and $\chi_S(r, t)$ for thresholds $d = a/2$, $a$, and $2 a$, where $a$ is the mean-square displacement over 1 ps.}
\end{center}
\end{figure}

One can visualize the structural relaxation time by investigating the decay of the first peak of the quantities $G(r, t)$, $\chi_N (r, t)$ and $\chi_S (r, t)$; this is shown in Fig.~\ref{fig:coherence_fp}. The decay of $\chi_N (r, t)$ and $\chi_S (r, t)$ for $d = a$ closely follows that of the Van Hove function, as we have already remarked above in our discussion of Fig.~\ref{fig:coherence_all}. For $d = 2a$ the decay of the four-point functions is more sluggish; this is a manifestation of the mantra that large objects evolve more slowly. In contrast, for $d = a/2$ the relaxation time of the four-point functions is substantially shorter, as a displacement of $a/2$ is typical in the phonon regime, where motion is ballistic and not caged, and the relaxation time is short. At long times ($t \gtrsim 8$ ps) the decay time constant of the four-point functions equals $\tau = 2.77$, 3.62, and 7.88 ps, for $d = a/2$, $a$, and $2 a$.

\section{Discussions and Concluding Remarks}
\label{sec:4}

We have seen that the four-point functions $\chi_S (r, t)$ and $\chi_N (r, t)$ -- defined by thresholding the atomic displacements of each atom $i$ or all atoms within some distance from it -- behave fairly similarly. Importantly, while the relaxation time of these four point functions increases with increasing threshold distance $d$, they carry the same information about medium-range order as the Van Hove function $G(r, t)$ and the pair-distribution function $g(r)$ do. In particular, if one chooses the threshold distance $d = a$, the mean-square displacement over 1 ps, then the four-point functions are almost equal to $G(r, t)$. 

The four-point functions with a window function were introduced in order to exclude small local fluctuations in atomic position and focus on coarse-grained density fluctuations. However, the MRO already represents such coarse-grained density fluctuations \cite{egami_2020a}. The first peak in $g(r)$ or $G(r, t)$ depicts atomic positions of near-neighbor atoms. In contrast, peaks beyond the first peak cover many atomic distances, and depict coarse-grained density correlations. In other words, they describe the point-to-set correlations, not the point-to-point correlations \cite{berthier_2012}. For this reason, the MRO portions of the PDF and the Van Hove function are already similar to the four-point functions in practice. This observation prompts us to identify medium-range order as the origin of dynamical heterogeneities.

Because the Van Hove function is much more easily determined in experiments than the four-point functions, and is also easier to interpret physically, it appears that the former has significantly greater utility than the latter. Indeed, since $g(r)$ already contains information about medium-range order and cooperative motion \cite{ryu_2021,egami_2021,lieou_2022a}, we posit that the four-point functions may not be needed, after all, to describe dynamical heterogeneities, structural relaxation, and deformation in glass-forming liquids. However, the four-point functions remain an interesting pedagogical tool that enables us to interpret the Van Hove function.

In an earlier paper \cite{wu_2018} it was argued that the de Gennes narrowing phenomenon, or the characteristic slowing down of dynamics near the maximum scattering intensity, has a geometric origin and does not necessarily imply dynamic cooperativity. Even for high-temperature liquids in which dynamic correlations are negligible, the relaxation time of the Van Hove function increases linearly with distance, resulting in apparent de Gennes narrowing. Thus, part of our present observation that the relaxation time scale of the four-point function depends on the size of atomic displacements may simply be related to geometrical de Gennes narrowing. When a size-dependent relaxation time is observed for a key quantitative tool commonly used to describe dynamical heterogeneities, one should not attribute it to dynamical heterogeneities without careful examination of the size dependence.

\begin{acknowledgments}
This work was supported by the US Department of Energy, Office of Science, Basic Energy Sciences, Materials Sciences and Engineering Division.
\end{acknowledgments}

\bibliography{pre_chi4_02}

\end{document}